\documentclass{article}
\usepackage{graphicx}
\usepackage{authblk}
\usepackage{xspace}
\title{IRIS-HEP 2026 Statistical Ecosystem Blueprint White Paper}
\date{March 2026}

\def\evermore   {\mbox{\texttt{evermore}}\xspace}
\def\everwillow {\mbox{\texttt{everwillow}}\xspace}
\def\Equinox {\mbox{\texttt{Equinox}}\xspace}

\def\HistFitter   {\mbox{\texttt{HistFitter}}\xspace}
\def\HEPData {\mbox{\texttt{HEPData}}\xspace}
\def\HST {\mbox{$\mathrm{HS}^3$}\xspace}

\def\JAX   {\mbox{\texttt{JAX}}\xspace}

\def\NumPy   {\mbox{\texttt{NumPy}}\xspace}
\def\Optax   {\mbox{\texttt{Optax}}\xspace}
\def\Optimistix   {\mbox{\texttt{Optimistix}}\xspace}

\def\pyhf   {\mbox{\texttt{pyhf}}\xspace}
\def\pyhs {\mbox{\texttt{pyHS3}}\xspace}
\def\paramore {\mbox{\texttt{paramore}}\xspace}

\def\RooFit {\mbox{\texttt{RooFit}}\xspace}
\def\RooStats {\mbox{\texttt{RooStats}}\xspace}

\def\Uproot   {\mbox{\texttt{Uproot}}\xspace}

\def\TRExFitter   {\mbox{\texttt{TRExFitter}}\xspace}

\def\zfit   {\mbox{\texttt{zfit}}\xspace}
\def\JSON   {\mbox{\texttt{JSON}}\xspace}

\def\CMSCombine   {\mbox{\texttt{CMS Combine}}\xspace}

\def\ROOT       {\mbox{\texttt{ROOT}}\xspace}

\def\Matplotlib {\mbox{\texttt{Matplotlib}}\xspace}

\def\cpp {\mbox{C\texttt{++}}\xspace}

\usepackage{jheppub}
\usepackage{hyperref}
\usepackage[titletoc]{appendix}
\usepackage{enumitem}
\setlist{noitemsep}
\usepackage[T1]{fontenc} % if needed
\usepackage[style=numeric-comp, sorting=none]{biblatex}

\begin{document}

\author[1]{\small Matthew Feickert}
\author[2]{\small Massimiliano Galli}

\affiliation[1]{\footnotesize University of Wisconsin--Madison}
\affiliation[2]{\footnotesize Princeton University}

\abstract{
This white paper presents the current status of the ecosystem of statistical tools used in High Energy Physics (HEP), and attempts to summarize the views on the future R\&D directions.
These views have been collected during the latest IRIS-HEP statistical ecosystem blueprint, which took place in February 2026.
}

\maketitle

\newpage
\section{Introduction}

%TODO
%Message: vast majority still uses Roofit/Roostats, but raising usage of ML-libraries (like JAX). We want to summarize the status and identify future hot topics and what to focus on.

%Maybe also elaborate on the separation of concerns (building, minimization, inference).

%This paper is organized as follows: ......

%The Institute for Research and Innovation in Software for High Energy Physics (IRIS-HEP)~\cite{IRISHEPSPUPDATE} blueprints are designed to inform the development and evolution of the strategic vision.
%The IRIS-HEP statistical ecosystem blueprint took place at CERN on February 24-25 2026.
%Mostly focused on the Large Hadron Collider (LHC) community, the blueprint had several goals: inspecting the state of the shared statistical ecosystem across the HEP community; identify and inspect possible gaps relevant for the High Lumi LHC; set a pathway to increase interoperability between existing tools (\ROOT) and Python based libraries; inspect and summarize approaches to bridgning the statistical and machine learning worlds.

%The participants included both developers and enthusiasts of the pythonic ecosystem and developers from the \ROOT team.
%The two-days blueprint covered different aspects: the first was mostly devoted to statistical model building, serialization and publication, while the second was mostly devoted to tools for minimization, inference.

The Institute for Research and Innovation in Software for High Energy Physics (IRIS-HEP)~\cite{IRISHEPSPUPDATE} organizes blueprint workshops to inform the development and evolution of its strategic vision.
The IRIS-HEP Statistical Ecosystem Blueprint took place at CERN on February 24-25, 2026, bringing together developers of the \ROOT-based and Python-based statistical tooling ecosystems.
The workshop had four main goals: assess the state of statistical tools across the LHC community; identify gaps relevant for the High-Luminosity LHC (HL-LHC); chart a path toward greater interoperability between \ROOT-based and Python-based libraries; examine the growing intersection between statistical inference and machine learning.

The first day focused on statistical model building, serialization, and publication, while the second addressed tools for minimization and inference.
This document summarizes the topics presented and discussions that took place and provides context for the ongoing developments.

The paper is organized as follows.
Section~\ref{sec:models} introduces statistical models in HEP, reviews open-world and closed-world approaches to model building, and describes the HEP Statistics Serialization Standard (\HST)~\cite{hs3} together with its funded development program.
Section~\ref{sec:ad} covers automatic differentiation in \RooFit through the Clad-based code generation pipeline and its impact on minimization performance.
Section~\ref{sec:jax} presents the emerging \JAX-based ecosystem, including \pyhf~\cite{pyhf_joss}, \evermore~\cite{evermore}, \pyhs~\cite{pyhs3}, and the \everwillow~\cite{everwillow} inference framework.
Section~\ref{sec:experiments} describes experiment-specific workflows: the \CMSCombine tool and the ATLAS tooling landscape.
Section~\ref{sec:histfactory2} introduces HistFactory v2, which replaces the traditional interpolation assumptions with Gaussian process regression.
Section~\ref{sec:statsml} discusses the intersection of statistics and machine learning through PHYSTAT and the VERAIPHY initiative.
Section~\ref{sec:future} outlines future prospects, and Section~\ref{sec:summary} provides a summary.

%\section{Scope of the Blueprint}

\section{Statistical Models, Interfaces, Serialization and Publication}
\label{sec:models}

%TODO: Introduction with stuff from~\cite{Cranmer:2021urp} about what a statistical model is and the importance of agreement of serialization and publication standards.

Experimental results in particle physics are based on statistical models $p(x, y | \mu, \theta)$, which define the probability for any set of observations given specific values of the input parameters.
Here $x$ denotes the primary measurement data, $y$ the auxiliary data constraining systematic uncertainties, $\mu$ the parameters of interest (such as a signal cross section or particle mass), and $\theta$ the nuisance parameters (calibration constants, background normalizations, and similar).
Inserting the observed data into this model yields the likelihood $L(\mu,\theta)$ from which all inference steps follow.
%The statistical model carries strictly more information than the likelihood: it retains the full dependence on both data and parameters, enables pseudo-data generation for frequentist procedures, and exposes individual signal and background components needed for combinations and reinterpretations~\cite{Cranmer:2021urp}.

The models built in HEP can be classified into two broad categories: open-world and closed-world~\cite{Cranmer:2021urp}.
Open-world models allow analysts to define arbitrary custom components while closed-world models restrict the analyst to a finite, well-documented set of building blocks and composition rules.
Status and challenges differ between open- and closed-world models when it comes to model building, serialization and publication.

%The importance of publishing not only the likelihood but the full statistical model, for the purpose of reproducibility, has been stressed in several contexts~\cite{Cranmer:2021urp}.
%Nowadays, the landscape of models publication is as follows.

Open-world models have relied on the \texttt{RooWorkspace} for serialization.
Workspaces capture the full model in a \ROOT-based container, but loading them requires access to the same custom libraries used at creation time, tying publication to a specific software stack.
Examples of this approach can be seen in \CMSCombine and will be explored later.

Closed-world models lend themselves to declarative specifications that decouple the model from any particular implementation.
For instance, in the case of HistFactory-based analyses, the \pyhf package introduced a \JSON format that encodes model and data in a single, \ROOT-independent file, and the ATLAS collaboration has published a growing number of full statistical models on \HEPData~\cite{Maguire:2017ypu} in this format. 
This \JSON approach improves portability and long-term archival, but covers only the HistFactory model class and does not extend to unbinned or custom-component analyses. 

The HEP Statistics Serialization Standard (\HST), discussed extensively during the blueprint, aims to bridge this gap by providing a single declarative standard broad enough to accommodate both HistFactory-class and more general statistical models, defining a \JSON-based format for serialization.
\HST is a standard, not an implementation: it specifies what a valid model description must contain and how each component maps to a mathematical definition, leaving framework developers free to choose their own realization.
Some frameworks already target \HST compliance: 
\RooFit/\RooStats covers nearly 100\% of the documented standard in \cpp.
\pyhs~\cite{pyhs3}, a pure-Python implementation, is under active development.

From a technical point of view, \HST adopts statistical language rather than HEP-specific jargon, aiming to make models accessible to statisticians and researchers outside particle physics.
The standard represents a statistical model as a directed acyclic computational graph, with named string identifiers linking objects for human readability.
An \HST document organizes the model into top-level components: distributions, auxiliary functions, datasets (binned or unbinned), domains (allowed parameter and observable ranges), parameter points (starting values, best-fit points, hypothesis settings), likelihoods (pairings of distributions and datasets), analyses (high-level entry points identifying a likelihood, domain, and parameters of interest), and metadata (provenance, authoring software, \HST version).

The standard supports both low-level and high-level model descriptions. 
At the low level, every distribution and function appears explicitly. This approach can easily produce $\mathcal{O}(100\mathrm{k})$ lines of JSON for real-world models.
At the high level, \HST provides compact declarative constructions such as \texttt{histfactory\_dist}, which encodes a full HistFactory channel in a syntax close to the \pyhf \JSON format. 
Since the majority of LHC analyses use HistFactory-class models, this high-level construction covers a large fraction of use cases while remaining compact and human-editable. 
Real-world models will rarely be written by hand; \HST therefore envisions a tooling ecosystem around programmatic model generation and model-to-model transformations such as combining likelihoods, correlating nuisance parameters, reparametrizing for EFT interpretations, and pruning systematics. 
% add this later?
%Identifying the minimal set of operations needed to make HS3 models usable by outside researchers — particularly theorists performing reinterpretations — remains an open question raised during the workshop discussion.

Two funded projects will drive \HST development over the coming years.
A Deutsche Forschungsgemeinschaft (DFG) project ("Public Likelihood Combination") supports a proof-of-concept EFT interpretation of published \HST models, with the explicit goal of demonstrating end-to-end usability for theorists and catalyzing broader adoption.
The larger DEMOS initiative ("Democratizing Models," approximately 8 FTE over three years) targets adoption of \HST beyond HEP through documentation, tutorial workshops, community governance, validation pipelines, converters, and a large library of example models. 
Together, these efforts position \HST as the reference serialization standard for statistical models in HEP and adjacent fields, with major advancements expected within the next three years.

Despite this momentum, the blueprint discussions highlighted several challenges.
An ongoing round-trip validation effort, aimed at exporting \RooFit workspaces to \HST and re-importing them through \pyhs, has uncovered cases where the \ROOT exporter deviates from the evolving specification or omits information needed for full model reconstruction.
The specification itself contains ambiguities, including inconsistent naming conventions and unclear semantics for generic distributions. 
%A practical obstacle is the lack of moderate-sized test workspaces: available examples are either trivial or full-scale models with thousands of parameters, making incremental debugging difficult.
Tooling support across the ecosystem also remains incomplete: \pyhf is actively adding \HST import capabilities and preliminary work has demonstrated successful construction of \JAX computation graphs from \pyhs, but other tools do not yet support the standard.
Broader adoption raises additional open questions, such as how to validate that an exported \HST file faithfully reproduces the original model and how to standardize systematic uncertainty naming across experiments. 
The funded DEMOS and DFG projects provide the resources to address these challenges, but progress will depend on the community converging on validation procedures and contributing test models at realistic scales.

%\section{Model Serialization and Publication}

\section{Automatic Differentiation with Clad in RooFit}
\label{sec:ad}

One of the most recent topics in the \ROOT world, which was covered during the blueprint, is the implementation of automatic differentiation in \RooFit.

Most HEP analyses build and minimize likelihood functions using \RooFit~\cite{Verkerke:2003ir} and Minuit2~\cite{Hatlo:2004sga}.
Three main bottlenecks limit the performance of this minimization: likelihood evaluation, gradient computation, and linear algebra operations within Minuit2.
Of these, gradient computation dominates for models with many parameters.
Minuit2 computes gradients numerically by default, varying each parameter independently, so the cost scales linearly with the parameter count.
\RooFit partially mitigates this through caching of intermediate results, but the gradient remains the leading cost for models with hundreds of parameters, which is a regime already common in Run 3 and expected to become standard at the HL-LHC.

Reverse-mode automatic differentiation (AD) addresses this scaling problem directly.
In reverse mode, the gradient cost differs from the function evaluation cost by a constant factor, regardless of the number of parameters.
%ML frameworks such as TensorFlow and PyTorch provide AD, but do not cover the full range of HEP statistical workflows. 
\RooFit integrates a dedicated AD engine based on Clad~\cite{Singh:2023xgx}, a source-code-transformation tool implemented as a Clang compiler plugin.
The integration relies on a code generation (\textit{codegen}) pipeline. 
\RooFit first translates its computation graph into a standalone \cpp function encoding only the mathematical operations of the likelihood.
%For combined fits with multiple channels, the generated code assigns each channel a separate function called in a loop, ensuring that JIT compilation time and stack memory scale linearly with the number of channels. 
Clad then differentiates this \cpp code to produce the analytic gradient, which \RooFit wraps back into an object usable by Minuit2.
%The nominal likelihood evaluation still uses RooFit's vectorized CPU backend with caching; only the gradient relies on the generated code.
Users activate the full mechanism through a single flag in the fitting call.

\begin{figure}[htbp]
  \centering
  \includegraphics[width=0.48\textwidth]{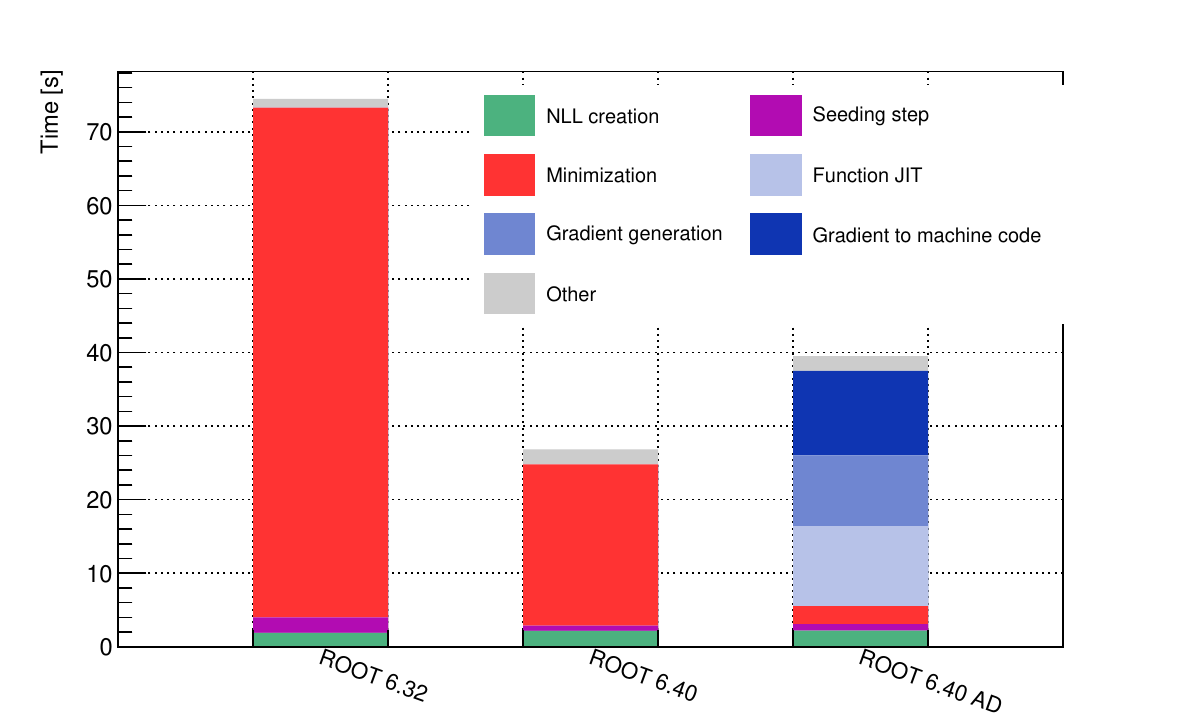}
  \hfill
  \includegraphics[width=0.48\textwidth]{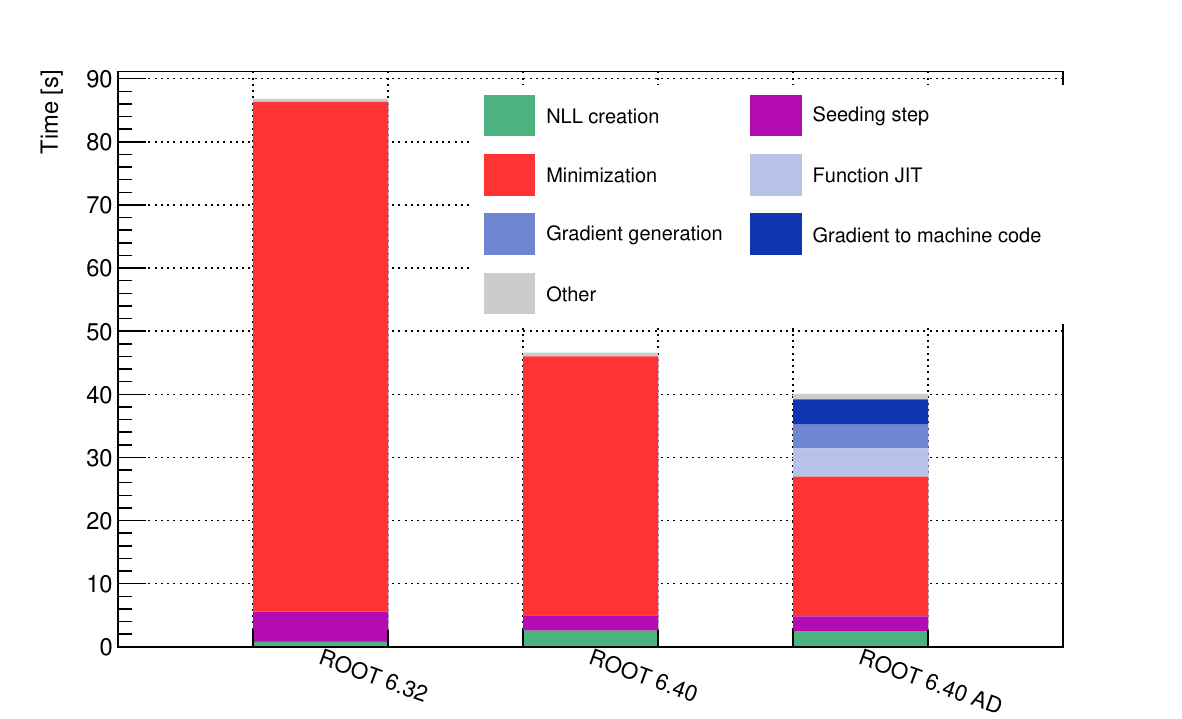}
  \caption{Fit performance comparisons for an ATLAS Higgs combination workspace (left) and CMS Higgs observation likelihood (right).}
  \label{fig:roofit_clad_studies}
\end{figure}

%TODO: Part about CMS and ATLAS studies with plots.
Examples of improvements in the overall minimization time are provided in Figure~\ref{fig:roofit_clad_studies}.
On the left, comparisons between \ROOT 6.32, \ROOT 6.40 and \ROOT 6.40 with AD are reported using an ATLAS HistFactory benchmark.
From this we can see that AD drastically reduces the minimization time (shown in red). The only caveat is the time taken to generate the gradient, JIT-compile the function and convert it to machine code. Since, however, this operation is performed only once, the main benefit comes from workflows that reuse this code (such as toy studies and likelihood scans).
On the right, similar comparisons are shown using the CMS Higgs observation model~\cite{cms_collaboration_2024_c2948-e8875}. The model in this case is more complex and heterogeneous, using 672 parameters and 102 channels with both template histogram and analytical shape fits. 
Also in this case, the red parts of the histograms show a decrease in the minimization time when using AD, although less drastic than in the other example. It should be noted though that \RooFit codegen support in CMS Combine is not optimized yet.

Several development directions follow from these results. 
The \RooFit team is working with CMS to deploy AD within \CMSCombine~\cite{CMS:2024onh}, targeting codegen coverage for all classes used in binned fits and support for analytical minimization of MC-statistic nuisance parameters within the generated code.
Analytic Hessian computation via Clad is on the \ROOT work plan for 2026, which would address the seeding-step bottleneck.
Further work aims at reducing JIT overhead and extending codegen to additional \RooFit primitives, including neural network components needed for simulation-based inference workflows.

\section{Toward a JAX-based ecosystem}
\label{sec:jax}

A growing interest is being dedicated to \JAX~\cite{jax2018github}.
\JAX is a \NumPy-compatible array library that extends standard array programming with composable functional transformations: just-in-time compilation (\texttt{jax.jit}), automatic differentiation (\texttt{jax.grad}), and explicit vectorization (\texttt{jax.vmap}). 
These transforms are composable and are applied at the level of a backend-agnostic intermediate representation (\texttt{Jaxpr}), before translation to compiler targets such as \texttt{XLA} and \texttt{LLVM}. 
This makes \JAX backend-agnostic by design and naturally suited to CPU, GPU, and TPU execution. 
\JAX integrates with the broader scientific Python ecosystem through the Array API standard, connecting to \Matplotlib, \NumPy, SciPy, \Uproot, boost-histogram, and iminuit, while also supporting a dedicated ML and optimization ecosystem including \Equinox~\cite{kidger2021equinox}, \Optax~\cite{deepmind2020jax}, and \Optimistix~\cite{optimistix2024}.

\begin{figure}[htbp]
  \centering
  \includegraphics[width=0.8\textwidth]{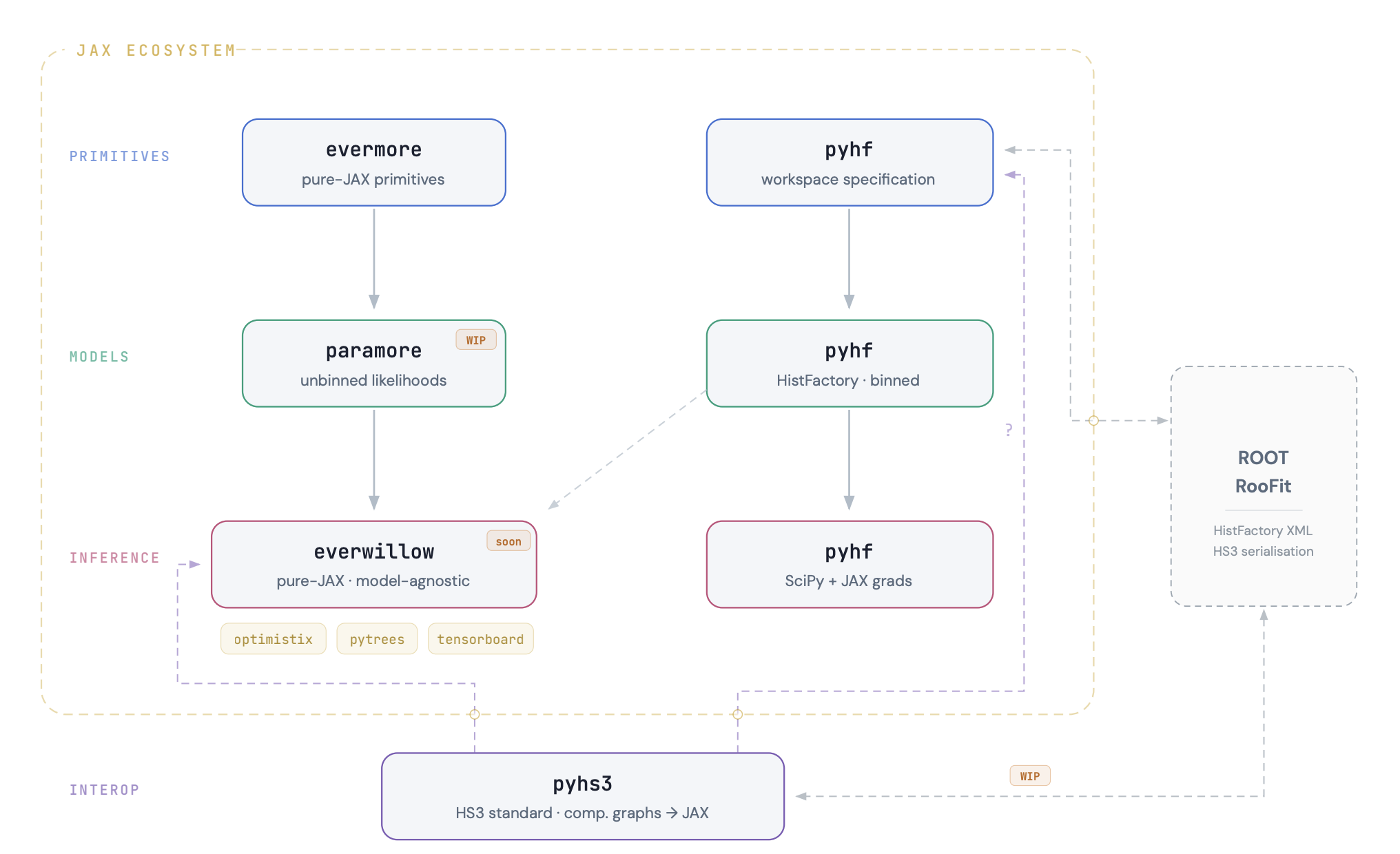}
  \caption{Sketch of the HEP \JAX-based ecosystem that is being developed.}
  \label{fig:jax}
\end{figure}

Several HEP-specific libraries are being developed on top of \JAX, forming a layered ecosystem organized around model specification, model building, inference, and interoperability.
A sketch of this ecosystem is reported in Figure~\ref{fig:jax}.
At the model specification level, \pyhf~\cite{pyhf_joss} implements HistFactory in Python with a \JAX backend and \JSON-serialized workspaces, enabling \texttt{jax.grad}, \texttt{jax.jit}, and \texttt{jax.vmap} to be applied directly to the likelihood.
\evermore~\cite{evermore} provides lower-level \JAX-native primitives using PyTrees (arbitrary nested Python containers that integrate natively with \JAX transforms), with parameters, modifiers, and effects as first-class objects. 
Custom modifiers reduce to lambda functions, and neural networks can be embedded directly as model components through compatibility with the NNX framework~\cite{flax2020github}.
At a higher level of abstraction, \paramore~\cite{paramore} (a work-in-progress package) extends \evermore to unbinned likelihoods and parametric models.
The \HST standard serves as the interoperability layer: \pyhs reads \HST \JSON workspaces and transpiles their PyTensor computation graphs into \JAX-callable functions, bridging the \ROOT and \JAX worlds.
%Existing RooFit models can reach JAX through the RooFit →\to
%→ HS3 →\to
%→ pyhs3 chain, and Combine workspaces follow the same path once HS3 exporters are in place.
It should be noted that, despite not being specifically presented or discussed during the blueprint, the second version of \zfit~\cite{Eschle:2019jmu} is being re-written in \JAX and would ideally also fit into this ecosystem, providing the tools to build both binned and unbinned likelihoods.

For inference, \everwillow~\cite{everwillow} is a likelihood-agnostic minimization tool built entirely in \JAX using \Optimistix, a pure-\JAX optimizer library supporting BFGS, L-BFGS, MIGRAD, SIMPLEX, and composable custom solvers.
Because optimizers in \Optimistix are themselves PyTrees, the entire minimization is end-to-end JIT-compilable. 
This enables applying \texttt{jax.vmap} over full minimizations, running hundreds of toy fits or parameter scans in a single vectorized call. 
\everwillow requires only that the likelihood be a \JAX function with parameters expressed as PyTrees; it supports bounded parameter transforms and provides full fit introspection via callbacks and mutable \texttt{jax.Ref} objects.
%Differentiation through the minimization itself is possible via the implicit function theorem, opening applications such as impact plot computation and NEOS-style \citep{neos} end-to-end analysis optimization.
Model combination across frameworks is handled by \texttt{statelib}, a utility within \everwillow that leverages PyTree flattening to merge parameter structures from different sources into a single unified state.
Each individual likelihood receives only its own parameter subtree, while the minimizer operates on the merged state. 
The combined negative log-likelihood (NLL) is simply the sum of the component terms.
%This approach requires no code changes in downstream libraries such as pyhf or zfit, as parameters are already expressed as PyTrees in those tools. 
Cross-model correlations between components built with different frameworks follow directly from the merged-state representation.

Another important outcome of the blueprint consisted in proving the interoperability of \JAX with \RooFit.
At the model level, we already mentioned how \HST could be the answer to interoperability between frameworks, provided that each framework implements the necessary machinery to import and export statistical models in \HST format.
However, while that is not in place, an interesting use-case for interoperability consists in performing combinations at the likelihood level, essentially adding negative log-likelihoods built in both \JAX and \RooFit and performing the minimization within either of the two worlds.
A prototype of this\footnote{\url{https://gist.github.com/guitargeek/ea90e36abb46928ff0c2e976b7b69c85}} was achieved during the blueprint, which demonstrates both directions of interoperability.
In the first direction, a \JAX-defined negative log-likelihood is wrapped into a \RooFit-compatible object through a small \cpp bridge class, allowing it to be summed with a native \RooFit NLL and minimized with Minuit2, with analytic gradients flowing from both sides.
In the second direction, a \RooFit NLL and its analytic gradient are exposed to \JAX through callback and custom differentiation mechanisms, making the \RooFit likelihood appear as a standard differentiable \JAX function that can be minimized with \Optimistix.
While still at the proof-of-concept stage, this prototype establishes that combining likelihoods across the two ecosystems is technically feasible and can preserve end-to-end differentiability.

Given its growing adoption across statistics and machine learning and its native support for automatic differentiation, JIT compilation, and hardware acceleration, \JAX is a natural candidate for broader adoption within the HEP statistical ecosystem.
Expanding the \JAX-based toolset and strengthening its interoperability with the established \RooFit ecosystem will be key exploration goals for the coming years.

\section{Experiment Specific Workflows}
\label{sec:experiments}

During the blueprint, the statistical inference workflows of the two main collaborations CMS and ATLAS were presented, and advantages and disadvantages of the two summarized and analyzed.

\subsection{CMS: Combine}

\CMSCombine~\cite{CMS:2024onh} serves as the primary statistical inference tool in CMS, used by approximately 92\% of analyses according to internal surveys.
It provides a command-line interface to \RooFit and \RooStats~\cite{Moneta:2010pm} methods, supporting counting, parametric (binned and unbinned), and template-based models.
Analysts define statistical models through human-readable datacards that specify the primary and auxiliary likelihood components.
A physics model step then translates these datacards into \RooFit workspaces.
\CMSCombine supports a range of statistical procedures recommended by the CMS Statistics Committee, including asymptotic limit setting~\cite{Cowan:2010js}, profile likelihood scans, goodness-of-fit tests, and nuisance parameter impact evaluation.

\CMSCombine implements several custom classes that optimize likelihood evaluation beyond standard \RooFit.
\texttt{CachingNLL} provides caching and constant-term optimizations for the negative log-likelihood, avoiding redundant PDF evaluations and supporting runtime channel masking.
For template-based fits, \texttt{CMSHistErrorPropagator} implements the Barlow–Beeston-lite technique~\cite{Barlow:1993dm}, analytically minimizing bin-by-bin MC statistical nuisance parameters rather than passing them to Minuit.
This reduces the effective parameter count and improves scaling.
\texttt{CMSHistSum} encapsulates all processes within a single object for vertical morphing of systematic variations, reducing memory usage by up to 60\% and minimization time by up to 40\% in large models. 
\CMSCombine also supports discrete profiling through a \texttt{CascadeMinimizer} that optimizes over combinations of discrete parameter indices.

Combinations are handled in \CMSCombine at the datacard level, before workspace creation. 
The most recent CMS Run~2 Higgs combination~\cite{CMS:2025jwz} illustrates the scale: it involves more than 1000 histograms, over 10000 total parameters, and over 90 parameters of interest, spanning template and parametric models across multiple decay channels.
Workspace creation required 26~GB of memory and the full scan suite took approximately two weeks on a batch system.

\CMSCombine is distributed through CMSSW, standalone Docker containers, and conda-forge. It has also been included in the ATLAS StatAnalysis stack, enabling cross-experiment use.

%\CMSCombine has begun publishing statistical models on HEPData, releasing Combine datacards alongside documentation and systematic uncertainty descriptions. 
On the development side, \CMSCombine is actively working on integrating the \RooFit AD codegen backend and on improving the user interface.

Since the publication and reinterpretability of the statistical models are goals of the collaboration, CMS has established a systematic pipeline for publishing statistical models alongside analysis results.
Analysts store their Combine datacards and \ROOT files in a central repository, which serves both analysis preservation and future combinations.
Each repository includes continuous integration tools that validate datacards, check naming conventions, build workspaces, and run standard statistical methods automatically.
Multiple review checkpoints ensure correctness before publication, and the process runs in parallel with \HEPData preparation to avoid delaying papers.
%CMS publishes its statistical models on the CERN Open Data repository (CDS) under a CC-BY-4.0 license, including the datacards, the Combine version and commands needed to reproduce results, a table of systematic uncertainties, and any auxiliary material. 
The entries published on the CERN Document Server (CDS) include the datacard, the \ROOT workspace, a table of systematic uncertainties, and the Combine version and commands needed to reproduce results.

%\HEPData entries link to the statistical model when available. Combine also supports reinterpretation of published models by allowing users to redefine parameters of interest, freeze nuisance parameters, or apply a different physics model to the same datacards. The CMS strategy is that all new analyses relying on Combine datacards will publish their statistical model by default.
As already mentioned in Section~\ref{sec:models}, this way of publishing the model heavily relies on specific software being available to reproduce the results.
For this reason, \CMSCombine is also working on integrating support for \HST importers and exporters.

\subsection{ATLAS: a Diversified Tooling Landscape}

ATLAS uses a more distributed approach to statistical tooling, with of order ten tools coexisting.
The most widely used public tools are \TRExFitter (\cpp/\RooStats), \HistFitter (Python/\RooStats), and \pyhf (Python, \ROOT-independent), with \pyhf adoption growing steadily.
Most tools function as wrappers around HistFactory, preprocessing inputs, assembling models from text or XML configuration files, and producing diagnostic outputs such as pull plots, ranking plots, pre-fit and post-fit comparisons, and likelihood scans.

The most common use case in ATLAS is the binned profile-likelihood fit using a HistFactory model. 
Beyond this, ATLAS analyses employ several unfolding methods depending on the measurement: iterative Bayesian unfolding, \texttt{TUnfold} ($\chi^2$-based), fully Bayesian unfolding, and profile-likelihood unfolding. 
%Profile-likelihood unfolding reframes differential cross-section extraction as a HistFactory fit, with per-bin signal strengths serving as the unfolded results. 
Unbinned likelihoods, used in analyses such as H$\rightarrow\gamma\gamma$, top mass, and B-physics measurements, rely on dedicated \RooFit-based frameworks, most of which are not public.

For combinations, ATLAS supports merging HistFactory workspaces through multiple tools, mostly Higgs-specific (\texttt{quickFit}, \texttt{XmlAnaWSBuilder}, \texttt{quickStats}). 
%Combined workspaces can reach approximately 1~GB, with tens of parameters of interest and thousands of nuisance parameters. 
In the case of non-HistFactory combinations, the BLUE method~\cite{Nisius:2020jmf} is often used. 
%Covariance matrices play a central role in ATLAS results, both as outputs for differential cross-section measurements and as inputs for uncertainty breakdowns \citep{atlas-covmat}.

ATLAS has adopted \HST for likelihood publication, with several analyses already published on \HEPData~\cite{Maguire:2017ypu} in this format.
Additional analyses use the \pyhf \JSON format.
%TODO: add from Giordon's talk.

Looking ahead, ATLAS identifies several pressing needs: models will grow more complex at the HL-LHC, with EFT fits requiring many simultaneous parameters of interest and pseudo-experiments increasingly needed for limit setting.
Alternatives or improvements to Minuit, better Hessian computation methods, and a common serialization language across tools and experiments rank among the priorities. 
The increasing role of ML-based approaches, including simulation-based inference and unbinned unfolding, also raises questions about result preservation and standardized formats for these new techniques.

\section{HistFactory v2}
\label{sec:histfactory2}

A second version of the HistFactory model was presented during the blueprint.

The traditional HistFactory model estimates expected event yields as a function of nuisance parameters $\boldsymbol{\alpha}$ through two key assumptions.
First, the impact of systematic uncertainties factorizes: the total effect on the yield is a product of individual per-parameter effects,
\begin{equation}
\Delta_{scb}(\boldsymbol{\alpha}) = \frac{\nu_{scb}(\boldsymbol{\alpha})}{\nu_{scb}^{0}} = \prod_{p}^{N_\mathrm{syst}} \frac{\nu_{scb}(\alpha_p)}{\nu_{scb}^{0}}\,.
\end{equation}
Second, the dependence on each parameter $\alpha_p$ follows a fixed parametric ansatz (piecewise linear, piecewise exponential, or polynomial interpolation), fitted to simulations available only at the nominal point and $\pm\delta_{\alpha_p}$ variations.
These assumptions work well for many analyses, but they have known limitations.
The factorization assumption breaks down when multiple systematic sources interact, since even independently chosen parameters can produce correlated effects on the final likelihood.
The parametric ansatz, constrained by just three anchor points per parameter, lacks flexibility for more complex dependencies and offers no estimate of the interpolation uncertainty itself.

HistFactory v2 proposes to replace these two assumptions with Gaussian process (GP) regression.
A Gaussian process defines a distribution over functions, fully specified by a mean function $m(\boldsymbol{\alpha})$ and a covariance kernel $k(\boldsymbol{\alpha}, \boldsymbol{\alpha}')$:
\begin{equation}
f(\boldsymbol{\alpha}) \sim \mathcal{GP}\bigl(m(\boldsymbol{\alpha}),\, k(\boldsymbol{\alpha}, \boldsymbol{\alpha}')\bigr)\,.
\end{equation}
Given a set of simulated yield predictions at arbitrary points $\boldsymbol{\alpha}^{(i)}$ in the high-dimensional nuisance parameter space ($\boldsymbol{\alpha} \in \mathbb{R}^{100\text{--}1000}$), GP regression produces a posterior prediction for the yield ratio $\Delta_{scb}$ at any new point $\boldsymbol{\alpha}$, distributed as
\begin{equation}
p\bigl(\Delta'_{scb} \mid \boldsymbol{\alpha},\, A_s^\mathrm{sim},\, \Delta_{scb}(A_s^\mathrm{sim})\bigr) = \mathcal{N}\bigl(\bar{\Delta}'_{scb},\, \mathrm{cov}(\Delta'_{scb})\bigr)\,,
\end{equation}
where $A_s^\mathrm{sim} = \{\boldsymbol{\alpha}^{(i)}\}$ denotes the set of $N_\mathrm{sim}$ simulation anchor points.
Crucially, these anchor points do not need to lie along orthogonal axes. 
Joint parameter variations are naturally accommodated, removing the factorization assumption.
The posterior mean $\bar{\Delta}'_{scb}$ serves as the yield prediction, while the posterior covariance provides an uncertainty estimate on the interpolation.
The choice of prior mean and covariance kernel injects an inductive bias analogous to the interpolation ansatz in HistFactory v1.
With appropriate defaults (e.g.\ the standard HistFactory interpolation as the prior mean and an RBF kernel with a large length scale $\ell$),
\begin{equation}
k(\boldsymbol{\alpha}, \boldsymbol{\alpha}_*) = \sigma_f^2 \exp\!\left(-\frac{\|\boldsymbol{\alpha} - \boldsymbol{\alpha}_*\|^2}{2\ell^2}\right)\,,
\end{equation}
the GP model reverts to the traditional behavior when only the standard $\pm 1\sigma$ simulations are available.

An additional refinement uses derivative information.
Passing approximate gradients $\partial_{\alpha_p} \nu_{scb}(\boldsymbol{\alpha})/\nu_{scb}^{0}$ at each anchor point further constrains the GP model, reducing interpolation uncertainties and capturing complex, non-factorizable effects.
These gradients can come from the simulation model or from numerical differentiation.
Optionally, the GP covariance matrix can feed into additional per-bin nuisance parameters $\gamma_b$ that propagate interpolation uncertainty into the fit through a constraint term
\begin{equation}
f_{\chi=\gamma}(\gamma_b, \boldsymbol{\alpha}) = \mathrm{Gaus}\!\left(1.0 \mid \gamma_b,\, \sqrt{\mathrm{cov}(\Delta'_{scb})}\right)\,,
\end{equation}
providing a principled way to account for modeling uncertainty in the systematics treatment.

The HistFactory v2 model retains the standard HistFactory treatment for unconstrained parameters ($\eta$: normalization factors, shape factors) and for bin-level statistical uncertainties ($\xi$: MC stat, shape systematics).
GP regression applies specifically to the constrained systematic parameters $\boldsymbol{\alpha}$, which drive the normalization and correlated shape variations, building the complete model $\nu(\eta, \xi, \boldsymbol{\alpha})$ without additional factorization assumptions.

Serialization of HistFactory v2 models requires extending the current format to support simultaneous multi-parameter variations alongside the traditional one-at-a-time variations, as well as optional gradient information at each anchor point.
The developers aim to harmonize this extension with the \HST standard.
Open questions raised at the workshop include whether the specification should prescribe the GP hyperparameters (kernel choice, interpolation code) or leave these to the implementation.
On the implementation side, the plan targets both \pyhf and \ROOT, with the \pyhf support provided either as a new package or as a backward-compatible extension of the existing library.
A paper documenting the HistFactory v2 specification is in preparation.

\section{Statistics Meets Machine Learning}
\label{sec:statsml}

\subsection{PHYSTAT}
\label{subsec:phystat}

As mentioned in the introduction, one of the goals of the blueprint consisted in exploring the growing connection between particle physics, statistics, and machine learning.
The PHYSTAT workshop series has served since 2000 as the main forum connecting professional statisticians and particle physicists \cite{Prosper:2011zz}. 
Machine learning introduces challenges and opportunities that extend beyond traditional statistics, and the PHYSTAT community has identified the need to include machine learning experts as a third group of collaborators alongside physicists and statisticians. 
The intersection takes three broad forms: using statistical theory to ground ML methods, using ML to implement statistical inference workflows, and using ML to learn optimal summary statistics from high-dimensional data.
Following two dedicated PHYSTAT workshops in 2024 on simulation-based inference (SBI) and on the broader statistics-meets-ML landscape, the community launched VERAIPHY (Validation \& Evaluation for Robust AI in PHYsics). 
VERAIPHY is an interdisciplinary effort to produce an overview document covering key areas at this intersection, including uncertainty quantification, model robustness, simulation-based inference, generative models, explainability, and symmetry. 
A first preview of the work took place at the PHYSTAT ML II workshop at Nikhef in February 2026. Together, PHYSTAT and VERAIPHY reflect a recognition that the HEP community's relationship with machine learning today mirrors its relationship with statistics two decades ago: physicists are skilled practitioners who would benefit from deeper collaboration with subject-matter experts.

IRIS-HEP is highly interested in the topics covered by PHYSTAT and VERAIPHY and looks forward to continuing to collaborate with these efforts.

\subsection{Interoperability with Simulation Based Inference Workflows}
\label{subsec:sbi}

Despite not being directly addressed during the blueprint, SBI represents one of the main examples of the intersection between physics, statistics and machine learning.
The traditional statistical procedure in HEP analyses builds the likelihood from histograms or parametric models, but this approach requires compressing high-dimensional data into low-dimensional summary statistics, often at the cost of information loss.
The purpose of SBI methods is to bypass this limitation by using machine learning techniques to extract inference directly from simulated samples, learning likelihood ratios (classifiers or regressors), optimal summary statistics or posterior densities (generative models) without requiring an explicit likelihood computation \cite{Cranmer:2015bka,Brehmer:2018eca}.
This enables analyses to work with higher-dimensional observables and capture correlations that could be discarded by the traditional binned approach, leading to an increase in sensitivity in many cases. Analyses that use SBI techniques at different levels have already been published by ATLAS~\cite{ATLAS:2025clx,ATLAS:2024jry} and CMS~\cite{CMS:2024ksn,CMS:2025dpp}.

From the point of view of the statistical tooling ecosystem, one major challenge is infrastructural and consists in performing joint fits (combinations) of SBI-based results with traditional workflows.
Research and development in this direction is being performed both within the Pythonic ecosystem and in the \ROOT-based one.

\section{Future Prospects}
\label{sec:future}

The blueprint discussions highlighted several priorities for IRIS-HEP investment over the coming years.

The first priority concerns \HST as the reference standard for model serialization, publication, and cross-framework interoperability.
The standard still requires development, particularly in providing a practical API for interacting with the underlying \JSON files, which can grow to considerable size for real-world models.
However, the path forward is clear: dedicated funding through the DFG and DEMOS projects provides a solid basis of person-power for the next few years, and adoption by both \RooFit and the \JAX ecosystem ensures broad coverage.

The second priority involves the development and support of \JAX-based tools.
As discussed in Section~\ref{sec:jax}, \JAX offers automatic differentiation, JIT compilation, and seamless interoperability with machine learning frameworks out of the box.
IRIS-HEP should support \JAX-native tools such as \evermore and \everwillow, and explore adopting \JAX as a backend for existing tools like \pyhf.

The third and closely related priority is interoperability between the \RooFit and \JAX worlds.
\RooFit and \RooStats remain central to the statistical workflows of the major LHC experiments, while the \JAX ecosystem offers complementary strengths in performance and extensibility.
Bridging the two can proceed at two levels.
At the model level, \HST and tools like \pyhs already provide a path: a model built in \RooFit can be serialized to \HST and transpiled into a \JAX-callable function.
At the likelihood level, the goal is to combine non-trivial negative log-likelihoods built with both \RooFit and \JAX into a single minimization, allowing analysts to use their preferred library for model construction while relying on shared tools for fitting and inference.
Achieving this interoperability would lower the barrier between the two ecosystems and enable the community to leverage the strengths of each.

A fourth priority is the integration of the statistical ecosystem with machine learning workflows.
As discussed in Section~\ref{sec:statsml}, machine learning techniques such as simulation-based inference are already entering mainstream LHC analyses, and their role will grow at the HL-LHC.
Combining SBI-based results with traditional profile-likelihood fits in joint statistical procedures remains an open infrastructural challenge.
Research and development toward this integration is underway in both the \ROOT-based and Python-based ecosystems, and IRIS-HEP considers this a key area of investment for the coming years.

\section{Summary}
\label{sec:summary}

This white paper has presented the current status of the statistical inference ecosystem in HEP and outlined the priorities identified during the IRIS-HEP Statistical Ecosystem Blueprint held at CERN in February 2026.

The \RooFit and \RooStats stack remains the backbone of statistical analyses at the LHC, and recent developments are extending its capabilities significantly.
Automatic differentiation through the Clad-based codegen pipeline delivers substantial speedups in minimization, with concrete benefits already demonstrated on ATLAS and CMS benchmarks.
In parallel, a \JAX-based ecosystem is emerging, built around \pyhf, \evermore, \pyhs, and the \everwillow inference framework, offering native support for differentiation, JIT compilation, and hardware acceleration.

Serialization, publication, and interoperability emerged as central themes. The \HST standard provides a path to a common, framework-independent description of statistical models, with funded development through the DFG and DEMOS projects.
%Validation efforts and tooling support are progressing but still face open challenges.
At the experiment level, CMS has established a systematic pipeline for publishing \CMSCombine-based models, while ATLAS continues to publish HistFactory and \HST models on \HEPData through its diversified tooling landscape.
HistFactory v2, based on Gaussian process regression, will extend the modeling capabilities of HistFactory-class analyses and is being designed to harmonize with \HST.

The blueprint also identified bridging the \RooFit and \JAX ecosystems as a key direction, with a first prototype of bidirectional likelihood interoperability demonstrated during the workshop. Finally, the growing intersection between statistics and machine learning, fostered by the PHYSTAT series and the VERAIPHY initiative, will shape the methodological landscape of HEP analyses in the coming years.

Taken together, these developments point to an ecosystem where consolidation around shared standards and increased interoperability between established and emerging tools will be central to meeting the demands of the HL-LHC era.

\section{Acknowledgments}

Matthew Feickert and Massimiliano Galli are supported by the U.S. National Science Foundation (NSF) under Cooperative Agreement OAC-1836650 (IRIS-HEP).

\sloppy
\raggedright
\clearpage
\printbibliography[title={References},heading=bibintoc]

\end{document}